\documentclass{aa}
\usepackage{graphicx}
\usepackage{txfonts}

\def\eq#1{{Eq.~(\ref{#1})}}

\begin{document}

\title{Constraints on supermassive black hole binaries from JWST and NANOGrav}
\titlerunning{Dual AGN from JWST and NANOGrav}

 \author{Hamsa Padmanabhan
          \inst{1}
          \and
          Abraham Loeb\inst{2}
          }

   \institute{D\'epartement de Physique Th\'eorique, Universit\'e de Gen\`eve \\
24 quai Ernest-Ansermet, CH 1211 Gen\`eve 4, Switzerland
   \\
              \email{hamsa.padmanabhan@unige.ch}
         \and
            Astronomy department, Harvard University \\
60 Garden Street, Cambridge, MA 02138, USA \\
             \email{aloeb@cfa.harvard.edu}            
             }

   \date{}

  \abstract
{We use the recent statistics of dual active galactic nuclei (AGN) in the \textit{James Webb Space Telescope} (JWST) data at $z \sim 3.4$ to address two aspects of the feedback and evolution scenarios of supermassive black hole binaries (SMBHBs). We find that the JWST data provide evidence for the members of a binary BH being `lit'  at the same time, rather than independently -- a scenario which is consistent with gas-rich mergers being responsible for concurrent AGN activity.  This conclusion is supported by the recent NANOGrav Pulsar Timing Array (PTA) measurements, whose  upper limits on the stochastic gravitational wave strain amplitude lie below those expected from extrapolating the dual AGN fraction. The results  indicate either a `stalling' of the binaries at the separations probed by NANOGrav, or rapid gas-driven inspirals.
}
 

\maketitle

\section{Introduction}

Galaxy mergers are expected to trigger accretion of gas into central supermassive black holes (SMBHs) of galaxies, forming pairs of active galactic nuclei (AGN). Studying the fraction of AGN in pairs, known as dual or binary AGN, can thus shed light on SMBH evolution.  It is currently unclear whether the mergers of SMBHs trigger AGN activity, leading to the simultaneous `shining' of the BH as dual AGN  \citep{Urrutia2008, Koss2010, Ellison2011, Schawinski2012, Treister2012, Satyapal2014, Villforth2014, Glikman2015, Glikman2018, Fan2016, Weston2017, Barrows2018, Goulding2018, Onoue2018, kocevskietal15,kossetal2018, Silverman_et_al_2011,Cisternas:2011,Mechtley:2016}, as opposed to `offset' AGN in which each BH shines separately. 
If SMBHs are not powered simultaneously by mergers \citep[e.g.][]{ventou2017, aird2019}, the fraction of dual AGN is expected to be below 2\% of all AGN.  From the few hundred dual AGN thus far identified in the literature, the studies of \citet{shen2023} and \citet{silverman2020} out to  $z \sim 4.5$ and $z > 1.5$  found dual fractions of 0.26 \% and 6.2 $\times 10^{-4} \%$, respectively. This has thus far been consistent with simulations \citep[e.g.][]{vanwassenhove2012} that predict a maximum of 4\% dual AGN candidates across $0 < z < 4$ (separated by scales of the order of a few parsecs to kiloparsecs).   

Recently,  the \textit{James Webb Space Telescope} (JWST) found a `surprisingly high' fraction of dual AGN in the high-redshift Universe at $z \sim 3-4$ \citep[][]{perna2023}.  This complements the  existing sample of dual and multiple AGN at high redshifts \citep[$z \sim 1-3$;][]{mannucci2022, ciurlo2023, perna2023a}. A BH pair of a slightly lower mass ($10^7 M_{\odot}$) has recently been discovered as an offset AGN by JWST \citep{ubler2023}.  Concurrently, the NANOGrav Pulsar Timing Array (PTA) 15 year and the European PTA results found 3$\sigma$ evidence for a stochastic gravitational wave (GW) background, likely originating from the mergers of supermassive massive black hole binaries (SMBHBs) over a range of redshifts \citep{epta2023,nanograv2023}.

In this Letter, we use the JWST dual AGN findings, in combination with analytical techniques and the recent NANOGrav PTA results to investigate two aspects of the evolution mechanisms of SMBHs\footnote{The relationship between the JWST SMBH and the PTA data was recently discussed in the context of primordial black holes in \citet{guo2023}.}: (i) the `final parsec' problem, in which binaries are stalled by the lack of stars of gas in their vicinity \citep[e.g.][]{milosavljevic2003} or the opposite effect -- that is, gas-rich mergers in which coalescence proceeds too rapidly to be detectable in low-frequency GW searches \citep[e.g.][]{kocsis2011}; and (ii) whether the two members of a dual AGN pair shine independently or at the same time, which, in turn, is related to the possibility of AGN being triggered by gas-rich  BH mergers \citep{treisteretal12,fan_etal2016,goulding+2018}.

\section{Relevant equations}

\begin{figure}
\begin{center}
\includegraphics[width =\columnwidth]{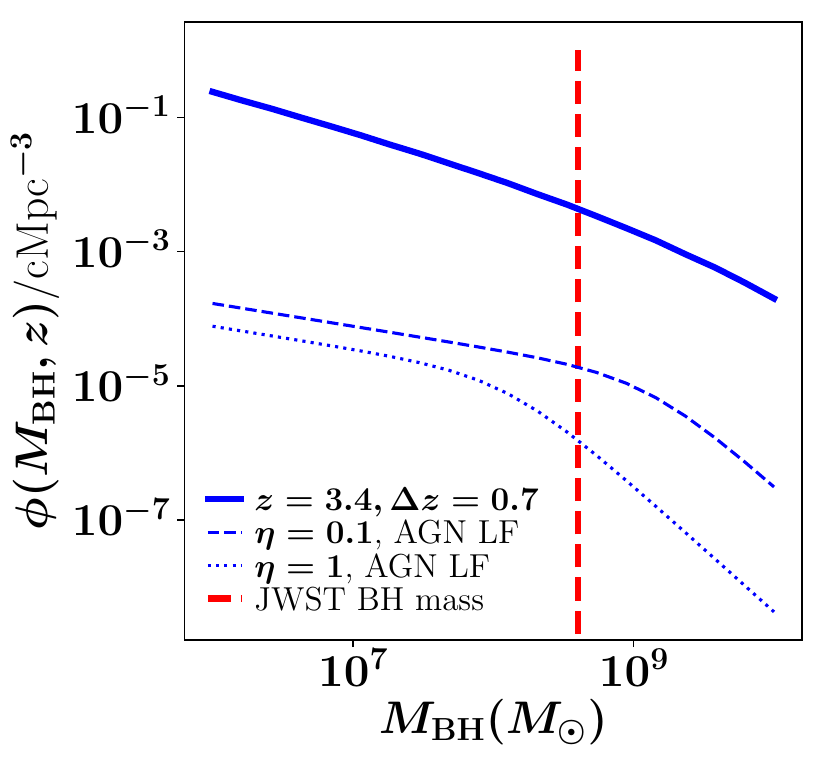}
\end{center}
\caption{Expected abundance  of BH mergers, evaluated using  \eq{phibhb} with $\Delta z = 0.7$, $q_{\rm min} = 0.1$, and $q_{\rm max} = 1$ (solid blue line). Also shown is the AGN bolometric luminosity function at $z \sim 3.4$ plotted for two different values of the Eddington ratio $\eta = 0.1$ and $\eta = 1$ (dashed and dotted blue lines, respectively). The fiducial BH mass probed by the JWST dual AGN sample is shown by the vertical dashed red line.}
\label{fig:lf_mergers}
\end{figure}

We use the formalism developed for the analysis of GW from mergers of SMBHBs \citep{hploeblisa2020, hploebpta}.  
It uses the merger rate of dark matter halos per unit 
redshift ($z$) and the halo mass fraction ($\xi$), as formulated by \citet{fakhouri2010},
\begin{eqnarray}
\frac{d n_{\rm halo}}{dz d\xi} &=& A \left(\frac{M_{\rm h}}{10^{12} 
M_{\odot}}\right)^{\alpha} \xi^{\beta} 
\exp\left[\left(\frac{\xi}{\bar{\xi}}\right)^{\gamma_1}\right] (1+ z)^{\eta}
\label{halomerger_rate}
,\end{eqnarray}
and it was fitted to the results of simulations over redshifts $z = 0-20$,  primary halo masses $M_{\rm h} = 10^{10} - 10^{15} M_{\odot}$, and mass ratios $10^{-5} < \xi < 1$.
The best-fitting parameter values are found to be $\alpha = 0.133$, $\beta = -1.995$,
$\gamma_1 = 0.263$, $\eta = 0.0993$, $A = 0.0104$, and $\bar{\xi} = 9.72 \times 
10^{-3}$.
To obtain the merger rates of BHs hosted by the dark matter haloes, we use 
the empirically constrained BH-halo mass relation \citep[e.g.][]{wyithe2002},
\begin{equation}
 M_{\rm BH} = M_{\rm h} \epsilon_0 \left(\frac{M_{\rm h}}{10^{12} M_{\odot}}\right)^{\gamma/3 - 
1} \left(\frac{\Delta_v \Omega_m h^2}{18 \pi^2}\right)^{\gamma/6} 
(1+z)^{\gamma/2} \, ,
\label{mbhmhalo}
\end{equation} 
in which $\log_{10} \epsilon_0 = -5.02 $ and $\gamma = 4.53 $ \citep[e.g.][]{hploeblisa2020}.  This relation is fitted to match the tight power-law correlation between the masses of BHs and their host galaxies in the local Universe \citep{ferrarese2002} and generalized to higher redshifts assuming a BH mass that scales with the circular velocity of the halo, $v_{c,0}$ as $M_{\rm BH} \propto v_{\rm c,0}^{\gamma}$ \citep{barkana2001, haehnelt1998}. This framework has been found to reproduce the data on the luminosity function of quasars \citep{wyithe2002, volonteri2003}.
Substitution of the BH-halo mass relation into the dark matter halo merger rate, \eq{halomerger_rate},
leads to the expected number of BH mergers as a function of redshift and mass ratio:
\begin{eqnarray}
&& \frac{d N_{\rm BH,mrg}}{dz dq}  = A_1 f_{\rm bh}
\left(\frac{M_{\rm h}}{10^{12} M_{\odot}} \right)^{\alpha} \nonumber \\
&& \  \times q^{3/\gamma - 1 +3\beta/\gamma} 
(1+z)^{\eta} \exp\left[\left(\frac{q}{\bar{q}}\right)^{3\gamma_1/\gamma}\right] 
 \label{bhmergerrate}
.\end{eqnarray}
Here,  $f_{\rm bh}$ denotes the fraction of dark matter haloes occupied by BHs; $q$ is the BH mass ratio and is related to $\xi$ as $q = \xi^{\gamma/3}$, $\bar{q}(\gamma) = 
\bar{\xi}^{\gamma/3}$, and $A_1 = (3/\gamma) A$. 
This  can be used to derive the number of BH mergers expected per unit comoving volume as
\begin{eqnarray}
&& \frac{d n_{\rm BH, mrg}}{dz dq d \log_{10} M_{\rm BH}} =  A_1 f_{\rm bh} \frac{3}{\gamma} 
\left(\frac{M_{\rm h}(M_{\rm BH})}{10^{12} M_{\odot}} \right)^{\alpha} \nonumber \\
&& \  \times q^{3/\gamma - 1 +3\beta/\gamma} 
(1+z)^{\eta} \exp\left[\left(\frac{q}{\bar{q}}\right)^{3\gamma_1/\gamma}\right] \frac{dn_{\rm h}}{d \log_{10} M_{\rm h}}
 \label{bhnumberdensity}
,\end{eqnarray}
\\
where $dn_{\rm h}/d \log_{10} M_h$  is the dark matter halo mass function [following, e.g. \citet{sheth2002}] and $M_{\rm h}(M_{\rm BH})$ denotes the dark matter halo mass as a function of BH mass, by inverting \eq{mbhmhalo}.
An integration over the relevant $q$ and $z$ ranges then gives 
the total number of SMBHBs expected per unit comoving volume and logarithmic mass interval:
\begin{eqnarray}
&& \phi_{\rm BHB} (M_{\rm BH}, z) \equiv \frac{dn_{\rm BHB}}{d \log_{10} M_{\rm BH} } \nonumber \\
&=& \int_{q_{\rm min}}^{q_{\rm max}} \int_{z_{\rm min}}^{z_{\rm max}}  dz \ dq \frac{d n_{\rm BH, mrg}}{dz \ dq \ d \log_{10} M_{\rm BH}}
\label{phibhb}
.\end{eqnarray}

If the separation between the BH pair is explicitly taken into account, the merger abundance is given by integrating \eq{bhnumberdensity} over the relevant time interval, converted to redshift
\begin{eqnarray}
&& \phi_{\rm BHB} (M_{\rm BH}, a) \nonumber \\
&=& \int_{q_{\rm min}}^{q_{\rm max}}  dq  \frac{dz}{dt} t_{\rm sep}(a) \frac{d n_{\rm BH, mrg}}{dz \ dq \ d \log_{10} M_{\rm BH}} \, ,
\label{phibhbasep}
\end{eqnarray}
where $t_{\rm sep}(a)$ is the binary decay time at separation $a$. For example, if the binaries are separated by a distance at which GWs are the main mode of decay, the separation time is given by $t_{\rm sep} (a_{\rm gr})$, which for BH masses $M_1$ and $M_2$ is given by 
the following \citep{peters1964}:
\begin{equation}
 t_{\rm sep}(a_{\rm gr}) = \frac{5}{256} \frac{c^5 a_{\rm gr}^4}{G^3 M_1 M_2 (M_1 + M_2)}
 \label{tgwagr}
.\end{equation}
The scale $a_{\rm gr}$ denotes  the radius at which the emission of GWs takes over as the dominant channel  of coalescence, defined by
\begin{equation}
a_{\rm gr} = A | {\rm ln} A|^{0.4} a_{\rm h}
.\end{equation}
Here,
\begin{equation}
A = 9.85 \left(\frac{M_1}{M_2}\right)^{0.2} \left(\frac{M_1 + M_2}{2 M_2}\right)^{0.4} \left(\frac{\sigma}{c}\right)
,\end{equation}
and
\begin{equation}
a_{\rm h} = 2.8 \left(\frac{M_2}{10^8 M_{\odot}}\right)  \left(\frac{\sigma}{200 {\rm km/s}}\right)^{-2} {\rm pc}
\end{equation}
is the hardening scale of the binary \citep[][]{merritt2000,kulkarni2012}.  In both expressions above, $\sigma$ denotes the velocity dispersion of the primary BH's host galaxy, related to its BH mass by \citep{tremaine2002}
\begin{equation}
\sigma = 208 \ {\rm km/s} \ \left(\frac{M_1}{1.56 \times 10^8 M_{\odot}}\right)^{1/4.02}
.\end{equation}
For the values of BH mass and separation considered in the following sections,  the timescale in \eq{tgwagr} is of the order of  $\sim 100$ Myr.

\section{Implications of recent data}
Next, we combine the formalism developed above with recent results from JWST to address two salient issues: (i) whether the two AGN in a binary or dual system are activated together or independently; and (ii) the possibility of a `final parsec' stall or fast, gas-rich mergers,  both of which decrease the strain observed by  recent low-frequency GW searches compared to expectations.

\begin{figure}
\begin{center}
\includegraphics[width =\columnwidth]{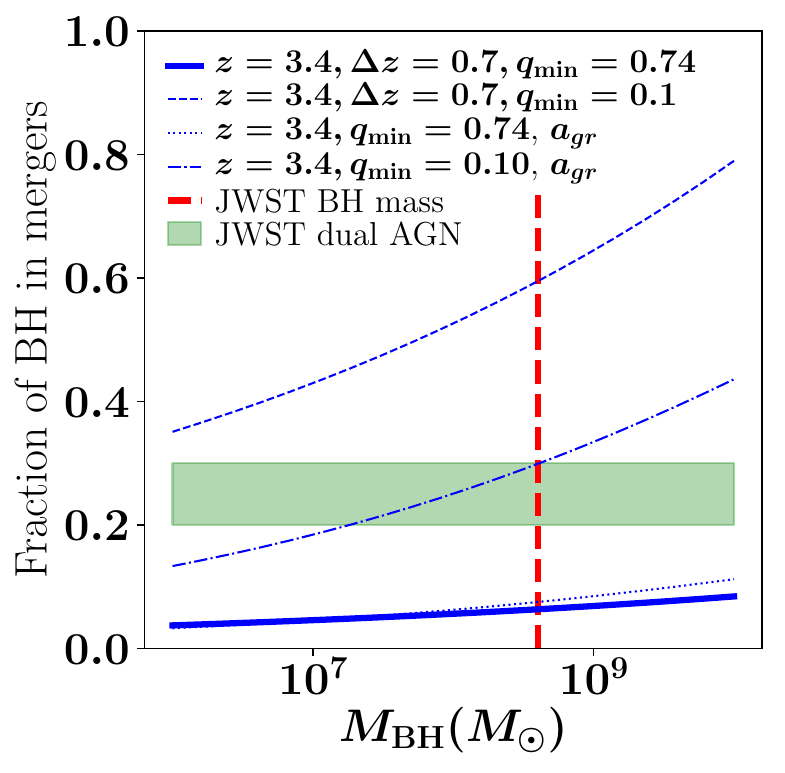}
\end{center}
\caption{Theoretically expected fraction of BHs in mergers, as a function of the primary BH mass at $z \sim 3.4$. The SMBHB abundances are evaluated for two values of the minimum mass ratio ($q_{\rm min} = 0.1, 0.74$), both independently of separation: \eq{phibhb} with $\Delta z = 0.7$ (dashed and solid blue lines) and at the GW decay-dominated separation, given by \eq{phibhbasep} with $a = a_{\rm gr}$ (dot-dashed and dotted blue lines). Overplotted as the green-shaded band  is the dual AGN fraction measured by JWST. The dashed red line shows the typical BH mass of the JWST sample.}
\label{fig:mbhratio}
\end{figure}

\subsection{Results from JWST}
\label{sec:sample}
{ The \textit{James Webb Space Telescope} reported the serendipitous discovery \citep{perna2023} of a triple AGN, three confirmed and one candidate dual AGN (centred at $z \sim 3.4$, covering the range $z \sim 3.1 - 3.7$). The AGN are located in the GOODS-S and COSMOS fields, and part of a larger 17 AGN sample at $2 < z < 6$ targeted by the JWST Galaxy Assembly with NIRSpec IFS (GA-NIFS) program. 
The \citet{perna2023} sample probes SMBHs with bolometric luminosities $10^{44.5} - 10^{46}$ erg/s, corresponding to a median BH mass of $\sim 10^8 M_{\odot}$  (and a range $M_{\rm BH} \sim 10^7 - 10^9 M_{\odot}$) for Eddington ratios of the order of 0.1 to 1.  We consider the AGN-A and AGN-B in addition to COS1638-A and COS1638-B as fiducial cases. In the latter pair, one of the SMBH has a measured BH mass of $M_{\rm BH}^{\rm fid} = (4 \pm 1) \times 10^8 M_{\odot}$,  which we use as a fiducial value. The stellar (galaxy) masses cover the range $10^8 - 10^{11} M_{\odot}$, and we use a value of $M_{\rm gal} \sim 10^{9.4} M_{\odot}$ ($\sim 10 M_{\rm BH}^{\rm fid}$) for the discussion here, corresponding to that of AGN-B.\footnote{ The stellar masses of the COS1638 AGN pair are both tentative estimates.  A large scatter in the measured stellar masses of some of the objects is reported in the literature \citep{weaver2022,jin2018}, due to photometric contamination and non-thermal emission precluding mass estimates based on the spectral energy distribution.} The separation between the SMBHs ranges from 4.7 - 28 kpc, with a median value $\sim 8$ kpc which we adopt here. Assuming the same value of $M_{\rm BH}/M_{\rm gal}$ for both  AGN of the dual systems, the ratio of BH masses is estimated to lie in the $q \gtrsim 0.63-0.8$ range for the fiducial cases (we adopt a median mass ratio $q \gtrsim 0.74$).  For completeness, we also quote results for a mass ratio $q \gtrsim 0.1$.}

The results suggest a 20-30\% dual AGN fraction in the population,   much higher than the maximum fraction of 1-5\% predicted by simulations \citep{vanwassenhove2012, maiolino2023, derosa2019} at these separations.

The expected dual AGN fraction can be calculated by using the AGN bolometric luminosity function measured at $z \sim 3.4$ \citep[from a comprehensive set of surveys covering $z \sim 0-7$;][]{shen2020}, and comparing it to the abundance of BH mergers, given by \eq{phibhb}.
These two quantities are plotted in  Fig. \ref{fig:lf_mergers} for two fiducial values of the Eddington ratio $\eta \equiv L/L_{\rm Edd}$, used to convert the bolometric luminosity of the AGN to the BH mass via $L_{\rm Edd}(M_{\rm BH}) = 1.3 \times 10^{38} M_{\rm BH}$. As can be seen, the number of shining BH at a given epoch is a couple of orders of magnitude lower than the number of BH mergers. This is expected because the merger abundance as calculated from \eq{halomerger_rate} is comparable to the BH abundance since there are order unity mergers per BH \citep{fakhouri2010}. { The ratio of AGN to all BH mergers is therefore comparable to the fraction of AGN to all BH. The latter is found to be close to the ratio of the typical AGN lifetime, $t_{\rm Q} \sim 100 \ $Myr, to the Hubble time, $f_{\rm active} = t_{\rm Q}/t_H$ \citep[e.g.][]{hploebpta}, which advocates that each halo merges, on average, roughly once per Hubble time \citep[e.g.][]{dong2022, genel2009}.}

\subsection{Do members of BH binaries shine independently or at the same time?}

It is currently unknown whether  AGN activity is triggered by mergers, with the variance in observations attributed to differences in sample selections and biases. If the dual AGN members are assumed to shine `together', or if the AGN activity is triggered by mergers, then the merging BH and dual AGN fractions should be comparable. On the other hand, if the AGN activity is non-simultaneous, then the dual AGN fraction should be much lower than the fraction of BHs in mergers, that is, less than $\sim 1$\% \citep{dorazio2023, vanwassenhove2012}.

We compare the theoretical predictions with the results from JWST in Fig. \ref{fig:mbhratio}. The fraction of BHs in mergers, computed independently of separation, is given by $\phi_{\rm BHB} (M_{\rm BH}, z)/\phi_{\rm BH} (M_{\rm BH},z)$, and that as a function of separation by $\phi_{\rm BHB} (M_{\rm BH}, a)/\phi_{\rm BH} (M_{\rm BH},z)$. In both of the above cases,  the BH mass function $\phi_{\rm BH} (M_{\rm BH},z)$ is defined as follows:
\begin{eqnarray}
\phi_{\rm BH} (M_{\rm BH},z) &\equiv& \frac{d n_{\rm BH}}{d \log_{10} M_{\rm BH}} \nonumber \\
&=& f_{\rm bh} \frac{d n_{\rm h}}{d \log_{10} M_{\rm h}} \left|\frac{d \log_{10} M_{\rm h}}{d \log_{10}  M_{\rm BH}}\right|
\label{blackholemfall}
.\end{eqnarray}
The fraction of BH in mergers, evaluated at $z \sim 3.4$ (both using \eq{phibhb} with  $\Delta z = 0.7$,  and \eq{phibhbasep} with $a = a_{\rm gr}$), each for two values of  $q_{\rm min} = 0.1, 0.74$,  and $q_{\rm max} = 1$, is plotted in Fig. \ref{fig:mbhratio}. The integral with $q_{\rm min} = 0.74$ corresponds to the JWST sample, while that with $q_{\rm min} = 0.1$ corresponds to the full $q$ range expected,  { since more extreme-mass ratio inspirals, with $q < 0.1$,  are  depleted due to the long dynamical friction timescales for small subhaloes in big haloes \citep{garrison2017, wetzel2010}.}

{ If the BH are assumed to proceed to $a_{\rm gr}$ without delay,  the merger fraction for both cases, $q_{\rm min} = 0.74$ and $q_{\rm min} = 0.1$ (dotted and dot-dashed blue lines, respectively) is comparable to, or lower than,  the JWST dual AGN fraction (shown by the green-shaded bar) for the masses within the range probed by JWST. This describes a scenario in which both AGN shine together, thus offering evidence in favour of AGN activity being triggered by mergers that proceed to coalescence via GW emission.  The fraction of BHs in mergers -- independently of separation -- over the redshift range probed by JWST is below the dual AGN fraction for $q_{\rm min} = 0.74$ (solid blue line), but it goes above it for $q_{\rm min} = 0.1$ (dashed blue line).  This implies that if the same fraction  (20-30\%) of dual AGN were to be found in an expanded sample covering the range $q_{\rm min} = 0.1$, the scenario of the AGN shining together would be disfavoured.}

\subsection{Evidence in favour of the final parsec problem or gas-rich mergers}

There is substantial uncertainty as to the evolution of SMBHBs after the hardening radius ($a_{\rm h}$). The BH may stall at this radius for $\sim 10^{10}$ yr or more (known as the final parsec problem) due to depletion of low-mass stars from the vicinity \citep{milosavljevic2003, merritt2000}. On the other hand, the final merging phase could be accelerated by the presence of gas or stars in the neighbourhood \citep{tiede2020, armitage2005, haiman2009}. 

We can use the previous results in combination with the measurements of the cumulative GW background from all binary BHs to address both possibilities. For this, we calculate the expected GW strain power spectrum for a  BH  population as a function of the observed frequency $f_{\rm obs}$, assuming only GW drive the evolution \citep[e.g.][]{phinney2001},
\begin{eqnarray}
&& h^2_{\rm c} (f_{\rm obs}) = \frac{4 \pi}{3 c^2} (2 \pi f_{\rm obs})^{-4/3} \int _{q_{\rm min}}^{q_{\rm max}}  \int_{z_{\rm min}}^{z_{\rm max}} \int_{M_{\rm min}}^{M_{\rm max}}  \nonumber \\
&& d \log_{10} M_{\rm BH} \  dz \  dq \frac{d n_{\rm BH, mrg}}{dz \ dq \ d \log_{10} M_{\rm BH}}  \frac{(G \mathcal{M}_{\rm ch})^{5/3}}{(1+z)^{1/3}} 
,\end{eqnarray}
in which ${d n_{\rm BH, mrg}}/{dz \ dq \ d \log_{10} M_{\rm BH}}$ follows \eq{phibhb} and the integral is performed over the BH population bracketed by a range of mass, $q$, and redshift. In the above equation, $\mathcal{M}_{\rm ch}$ denotes the chirp mass, which is defined by
\begin{equation}
\mathcal{M}_{\rm ch} = (M_1 M_2)^{3/5}/(M_1 + M_2)^{1/5}
\end{equation}
in the rest frame of the binary.

\begin{figure}
\begin{center}
\includegraphics[width =\columnwidth]{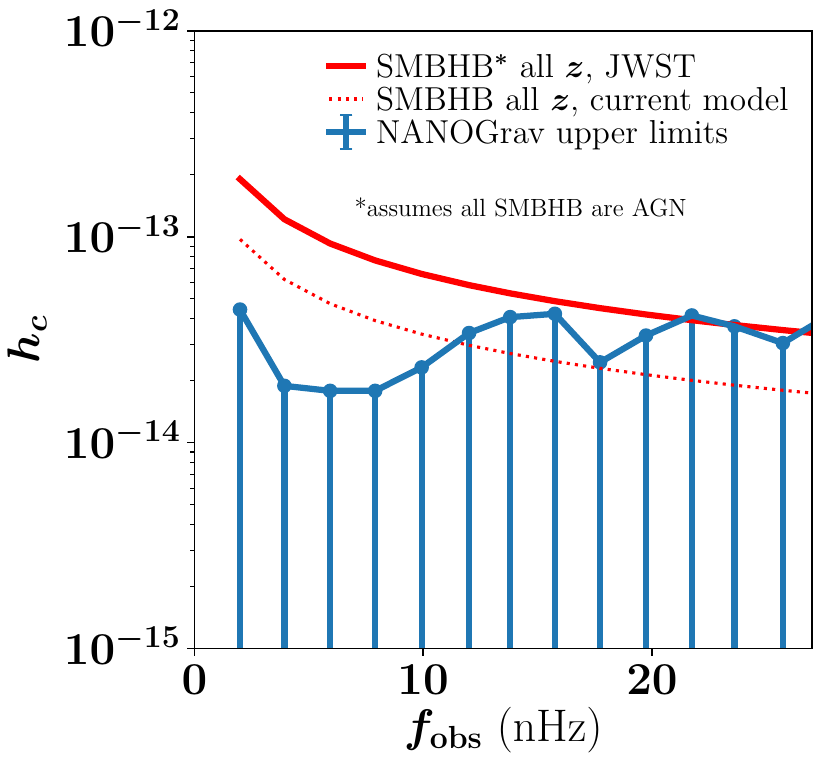}
\end{center}
\caption{Predicted strain versus frequency of the GW background with the fiducial model parameters, assuming circular binary evolution purely by GW emission (dotted red line), compared to the NANOGrav upper limits \citep{nanograv2023}. Also plotted is the expected strain implied by the JWST results, over all $z$, assuming dual AGN are (driven by) SMBHBs and proceed to coalescence via GW emission (solid red line).}
\label{fig:hcfrequency}
\end{figure}

The strain power spectrum calculated with the fiducial model parameters, integrated over the redshift range $\{0,7\}$, $q$ range $q_{\rm min} = 0.1, \ q_{\rm max} = 1$, and $M_{\rm BH} = \{10^6, 10^{10}\} M_{\odot}$, is shown in Fig. \ref{fig:hcfrequency} as the dotted red line.  Overplotted are the measured upper limits of the NANOGrav strain power spectrum \citep[][ solid blue error bars]{nanograv2023}. If the dual AGN fraction measured by JWST \citep{perna2023} is assumed to be representative of the general BH population, there is an increase in the predicted strain, given by the square root of the ratio of the median JWST AGN fraction (green-shaded band in Fig. \ref{fig:mbhratio}) to the solid (or dotted) blue line in that figure. The strain thus enhanced  is plotted as the solid red line in Fig. \ref{fig:hcfrequency}.  \footnote{The enhancement assumes that the {ratio} of the dual AGN fraction to that of the BH mergers is independent of the $q$ range under consideration, and hence the hike is uniformly applied over the whole range. This assumption can be refined with  larger samples of dual AGN in the future, covering lower $q_{\rm min}$ values.}

The gap between the predicted results and observed upper limits offers evidence for the depletion of the SMBHBs at the largest separations. This argues in favour of either (a) a final parsec-like problem causing the stalling of AGN, or (b) predominantly gas-rich mergers, which decrease the number of binaries observable by the PTA by shrinking the timescale needed for the BH to merge \citep{kocsis2011}. The latter scenario is further supported by the dual AGN shining together, enabled by the large reservoir of gas feeding both BHs.

{ We remark that the first possibility above, in which the BHs stall prior to coalescence, also has implications for the effect we considered in the previous section, namely the possibility of the AGN shining together. Specifically, if the binaries are assumed to stall before they reach the required distance for dominance of GW emission, the dynamical friction timescale can reach \citep{volonteri2020, binney2008}
\begin{equation}
t_{\rm df} = 0.67 \ {\mathrm{Gyr}}   \left(\frac{a}{4 \ {\rm kpc}}\right)^2 \left(\frac{M_{\rm BH}}{10^8 M_{\odot}}\right)^{-1}  \left(\frac{\sigma}{100 \ {\rm km/s}}\right) \frac{1}{\ln \Lambda}
,\end{equation}
where $\Lambda = 1 + M_{\rm gal}/M_{\rm BH}$ and $M_{\rm gal}$ is the (stellar) mass of the galaxy.
With $a \sim 8$ kpc,  $M_{\rm BH} = 4 \times 10^8 M_{\odot}$,  $\sigma = 200$ km/s, and $M_{\rm gal} = 10 \ M_{\rm BH}$ ({the fiducial values of the \citet{perna2023} sample adopted in Sec. \ref{sec:sample}}),  we find $t_{\rm df} = 0.71$ Gyr.
This, when used in place of $t_{\rm sep}(a_{\rm gr})$ in \eq{tgwagr}, corresponds to about a factor $\sim 2$ increase in the dotted curve in Fig. \ref{fig:mbhratio} (for $q_{\rm min} = 0.74$,  with $t_{\rm sep} (a_{\rm gr}) = 445$ Myr).  For the range of masses probed by the JWST observations, the scenario of AGN shining together is still favoured.  In the extreme case of all the BHs in the observable range being stalled by dynamical friction ($q_{\rm min} = 0.1,  t_{\rm sep} (a_{\rm gr}) = 68$ Myr),  this scenario may be disfavoured.
}

\section{Discussion}

We have used the recent JWST dual AGN statistics along with the measurements of the GW strain from the NANOGrav 15 year dataset to address two aspects of SMBH evolution. { If the SMBHs proceed to coalescence without delay, the fraction of dual AGN data measured by JWST is consistent with a scenario in which the AGN are triggered simultaneously by a SMBHB merger, leading to their concurrent activity.} Such a merger, if driven by a gas reservoir feeding both BHs, leads to a reduction in the signal observable with pulsar timing arrays, as compared to the case when the binary evolves purely by GW emission.

The probability of observing a binary BH as an AGN at a given separation $a$ scales as $N = t_{\rm sep}(a)/{t_Q}$, where $t_Q$ is the lifetime of the quasar \citep{foord2023, derosa2019}. The inverse of this quantity $N^{-1}$ gives the proportion of AGN concurrent with mergers at a given separation (which is, in turn, connected to the AGN being triggered by the merger). The duration of the merging phase when the BHs are separated by distances for which the GW emission is the dominant channel for decay, fed by the gas reservoir,  is comparable to the quasar lifetime, both found to be of the order of 100 Myr for the range of masses and mass ratios considered here.

Calculations similar to the ones in Fig. \ref{fig:mbhratio}  at $z \sim 0$ also lend support to the AGN shining together, where it is found that about 20-60\% of AGN with primary BH masses $\sim 10^9 M_{\odot}$ may be duals \citep{Koss2012}. This is, in turn, consistent with the paradigm of both AGN in the  pair being fed from the same gas reservoir. Given that the fraction of gas-rich, `wet' mergers increases with redshift,  these results are expected to hold out to higher redshifts as well.  { The bulk of the signal probed by NANOGrav is expected to arise from SMBHs at low redshifts $z \lesssim 1$ \citep{nanograv2023}. Noting that the BH merger rate is greater at $z \sim 3$ than $z \sim 1$, we expect that a similar dual AGN fraction ($\sim 20-30$\%) at $z \lesssim 1$ will imply a greater hike in the GW strain than predicted by the solid red curve of Fig. \ref{fig:hcfrequency}.}

There is some model dependence in the calculation of the amplitude of the GW strain. Some models based on BH-bulge mass observations \citep{bernardi2010} lie below the NANOGrav signal \citep[e.g.][]{sato2023}  at large separations. When the enhancement (by factors of a few) arising from the JWST dual AGN fraction is incorporated, most scenarios overproduce the signal compared to NANOGrav. The JWST data thus provide evidence for depletion of SMBHBs or a greater proportion of gas-rich mergers  -- which are believed to be responsible for AGN activity --  attenuating the PTA signal due to the rapid shrinking of their orbits \citep{kocsis2011, haiman2009, armitage2005}. Deviations from circular orbits, as assumed throughout the present calculations, may also arise from wet mergers \citep[e.g.][]{sobolenko2021}. Future PTA data, in combination with larger samples of dual and binary AGN expected with, for example, X-ray facilities \citep{foord2023}, will help narrow down the possibilities and improve our understanding of BH evolution and feedback mechanisms.

\section*{Acknowledgements}  
We thank A. Gopakumar for useful comments on the manuscript, { and the referee for a helpful report that improved the content and presentation.}
HP's research is supported by the Swiss National Science Foundation under Ambizione Grant PZ00P2\_179934. The research of AL is supported in part by the Black Hole Initiative,  which is funded by grants from the JTF and GBMF.

\def\aj{AJ}                   
\def\araa{ARA\&A}             
\def\apj{ApJ}                 
\def\apjl{ApJ}                
\def\apjs{ApJS}               
\def\ao{Appl.Optics}          
\def\apss{Ap\&SS}             
\def\aap{A\&A}                
\def\aapr{A\&A~Rev.}          
\def\aaps{A\&AS}              
\def\azh{AZh}                 
\def\baas{BAAS}
\def\jcap{JCAP}
\def\jrasc{JRASC}             
\def\memras{MmRAS}
\def\na{New Astronomy}
\def\nat{Nature}
\def\mnras{MNRAS}             
\def\pra{Phys.Rev.A}          
\def\prb{Phys.Rev.B}          
\def\prc{Phys.Rev.C}          
\def\prd{Phys.Rev.D}          
\def\prl{Phys.Rev.Lett}       
\def\pasp{PASP}               
\def\pasj{PASJ}
\def\physrep{Phys. Repts.}
\def\qjras{QJRAS}             
\def\skytel{S\&T}             
\def\solphys{Solar~Phys.}     
\def\sovast{Soviet~Ast.}      
\def\ssr{Space~Sci.Rev.}      
\def\zap{ZAp}                 
\let\astap=\aap
\let\apjlett=\apjl
\let\apjsupp=\apjs

\bibliographystyle{aa}
\bibliography{mybib, references, references_1}

\end{document}